# EnergyREV

# Cyber–physical components of an autonomous and scalable SLES

Nandor Verba, Pablo Baldivieso-Monasterios, Siyuan Dong, Andrei Braitor, George Konstantopoulos, Elena Gaura, Euan Morris, Alison Halford and Colin Stephen

December 2021

## Authors

- Nandor Verba | Coventry University
- Pablo Baldivieso-Monasterios | Sheffield University
- Siyuan Dong | Keele University
- Andrei Braitor | Sheffield University
- George Konstantopoulos | Sheffield University
- Elena Gaura | Coventry University
- Euan Morris | University of Strathclyde
- Alison Halford | Coventry University
- Colin Stephen | Coventry University

## Contents

# Executive summary

Adding renewable energy sources and storage units to an electric grid has led to a change in the way energy is generated and billed. This shift cannot be managed without a unified view of energy systems and their components. This unified view is captured within the idea of a Smart Local Energy System (SLES). Currently, various isolated control and market elements are proposed to resolve network constraints, demand side response and utility optimisation. They rely on topology estimations, forecasting and fault detection methods to complete their tasks. This disjointed design has led to most systems being capable of fulfilling only a single role or being resistant to change and extensions in functionality. By allocating roles, functional responsibilities and technical requirements to bounded systems a more unified view of energy systems can be achieved.

This is made possible by representing an energy system as a distributed peer-to-peer (P2P) environment where each individual demand energy resource (DER) on the consumer's side of the meter is responsible for their portion of the network and can facilitate trade with numerous entities including the grid. Advances in control engineering, markets and services such as forecasting, topology identification and cyber-security can enable such trading and communication to be done securely and robustly. They also provide users with improved metrics such as energy costs, $CO_2$ savings, grid constraints satisfaction and community-related utilities. Defining abstract roles and relations between components also aids in the development of agent and artificial intelligence (AI)-based approaches.

To enable this advantage however, we need to redefine how we view the design of the sub-systems and interconnections within smart local energy systems (SLES). In this paper we describe a way in which whole system design could be achieved by integrating control, markets and analytics into each system. We propose the use of physical, control, market and service layers to create 'system of systems' representation, where each sub-system is responsible for its respective area while receiving set points and references from its local group orchestrators. Understanding, representing, and optimising these cyber-physical interactions lies at the heart of designing and implementing effective SLES networks.

# Background and rationale

### Energy 3.0 – Energy system of the 2000's

The Energy 3.0 revolution was set up in a period of continuous growth and structured around an economic model of selling energy efficiently. In the early 2000s, a typical energy system had limited digitalisation and implemented fixed tariff relationships between the fairly static consumption and generation layers of the network. Digital services such as monitoring and control were encountered mainly at medium and high voltage levels. They were limited to load shedding and large energy resource management, with the end goal of satisfying network load power demand. The main beneficiaries of this level of digitisation were large scale energy producers who imposed fixed contracts, since their energy systems could then operate in the context of more predictable and easily controllable supply and demand cycles at a macro level. Moreover, high voltage energy systems were managed by simple control rooms that would mainly ensure the balance between generation and consumption in the grid. However, fixed tariffs meant static markets, with limited play in local pricing. Thus, both consumers and producers had little to no room to improve their overall costs. As a result, this energy system model needed rethinking in order to integrate additional constraints such as global emission, climate change and market expandability, and ways to move out of the financial and energy crisis.

### Energy 3.2 – Partially digitalised energy system

The current model, Energy 3.2, aims to develop along the lines above. It can extend further than Energy 3.0 because the emergence of centralised digital services, use-case tailored solutions and digital extensions to existing energy systems mean the transmission and distribution power system is not the only connecting part between generation and consumption anymore. Now, the system also incorporates additional control such as local optimisations and centralised network management, services focused on data capture, storage and reporting, and even energy markets with variable rates and prosumer contracts for small and medium producers. In addition, the emergence of the Internet of Things (IoT) in the energy sector contributed to the development of the so-called smart grid. This evolution towards adaptable and more distributed measurement and control for energy systems has been motivated by the increasing variability in energy supply from sources such as wind and gas. When variability grows, flexibility and adaptability are required to meet demand in a controlled way, to smooth out the highs and lows of supply and to provide consumers with a reliable energy source. This has led to the current Energy 3.2 framework. The benefits are vast, consisting of improved adaptability to multiple scenarios, better dynamic costs, and enhanced historical data storage and reporting. However, this model is still a single block system that lacks P2P trading and also lacks any consistent framework for fine-grained understanding or control of subcomponent responsibilities within such a many-layered system. Digital components aim to perform single tasks and are unable to guarantee reusability or upgradability, such as modifying current optimisers to predict future behaviour. In this increasingly complex setting of energy network component types and the relations between them, a digital revolution is needed to shape the future of electricity.

## Energy 4.0 – Fully digitalised

The evolution of digital technologies and the scaling of IoT deployment in all industry sectors has not only led to the new Industrial Revolution 4.0, but also to the early stages of Energy 4.0. In Energy 4.0, the energy system model incorporates fully digitalised energy resources that can interact dynamically and form local DER units that trade and optimise their functionality based on user needs and local measurements. In such systems, complex digital services aim to predict future behaviour, identify faults and optimise other key performance metrics.

The distributed and P2P energy markets provide flexibility, with demand-side response markets allowing individuals to choose how and which market to join. On the control side, swappable multi-objective local optimisers and controllers are introduced, supported by regional optimisers on top. If the model works well the result should be a fully customisable energy system tailored to consumer needs. The energy markets and controllers would achieve both local and national objectives: they would be interoperable and meet local demand. They would also satisfy national regulation and ensure consumer data privacy etc. But, with all its potential benefits, Energy 4.0 will lead to highly complex and heterogenous systems. Implementing Energy 4.0 principles while meeting practical, regulatory, and optimisation requirements requires a high computational cost. It also introduces an increased potential for security risks including cyber-attacks and privacy intrusions.

## Comparing the stages of the energy revolution

It is well accepted that an increasingly connected and intelligent ecosystem of energy assets would benefit from optimisation using big data and AI. Digital services evolved from being almost non-existent, in Energy 3.0, to being proposed as AI and machine learning (ML) driven, in the nascent Energy 4.0. This evolution not only presents potential for significant optimisation and reduction in electricity use, and therefore greenhouse gas emissions, but also offers options to use renewable generation en masse, and significantly improve the energy efficiency of existing assets and processes.

In order to realise the full benefits of Energy 4.0, the energy sector needs to address additional issues that arise during the transition to such complex systems. These include legislation and regulation as well as privacy risks. Debate and design must involve industry, government and the general public, to achieve inclusive sustainable solutions. Although there are a number of technical challenges to be overcome to implement in full Energy 4.0 principles, they will be outweighed by the benefits of this major digitalisation trend.

Table 1: Overview of energy systems evolutions when considering control, markets and digital services

| Use-Case | Control | Energy Markets | Digital Services |
|---|---|---|---|
| Energy 3.0 | Only medium and high voltage lines are managed. Main purpose is constraint satisfaction. | Fixed tariffs | Medium and high voltage line monitoring. Little to no public data availability |
| Energy 3.2 | Local optimisations. Centralised network management | More variety in tariffs and contracts (such as Demand Side Response). No peer-to-peer trading | Centralised digital services. Focused on low granularity and summary-level data capture |
| Energy 4.0 | Upgradable local optimisers and controllers. Regional community optimisers | Distributed peer-to-peer markets. Flexibility and demand-side response markets. Choice in markets, objectives and pricing | AI and ML based services. Identifying future behaviour, faults, metrics and key KPIs. Household level monitoring and optimisation |

Table 2: Overview of energy system evolutions benefits and drawbacks

| Use-case | Benefits | Drawbacks |
|---|---|---|
| Energy 3.0 | Robustness. Low volatility. Lower complexity | Inflexible pricing. Limited adaptability to varying consumption patterns |
| Energy 3.2 | Improved adaptability. Limited cost adaptability. Historic data analysis | Monolithic digital designs. Single purpose components. Lack of reusability |
| Energy 4.0 | Tailored, fully digitalised energy systems. Markets and controllers fulfil local and national objectives | High system complexity. Difficult to identify fraud. High costs for computing |

# The importance of scaling in SLES

Moving towards the integrated systems of Energy 4.0 is made easier by a key feature of SLES: self-similarity. This means that the similar network representation models can be applied to multiple layers within the energy network. For example, at the lowest layer, the interconnection of several renewable energy sources, storage devices, and local loads can be modelled in a similar way to an interconnected community, i.e., both share interconnection properties which allow energy flow across their components. Larger portions of the national grid can be conceptualised in the same way. At its very core, the fundamental principle all these networks share, regardless of their scale, is the ability to route and direct flows of energy, a concept that is not necessarily restricted to electric systems.

To build the integrated energy systems of the future we can use the above abstraction to further develop our understanding of a networked energy system; we can define operations and services that can be extrapolated from smaller scales to larger ones. The two main elements that we seek to explore here are control and market elements. These elements constitute a common feature among all scales of the network. However, these are not the only common features. Each scale is endowed with services corresponding natively to its scale. For example, one of the concerns of control elements at the smallest scale is to ensure that devices connected to renewable sources operate within safe margins; whereas in an upper layer, the control element focuses more on cooperation amongst network elements to guarantee constraint satisfaction at a network level. The market elements operate in a similar way: local areas are tasked to seek revenue for a community based on their generation and consumption, while for smaller scales the focus is on balancing consumption and generation. As the role of these control and markets elements changes with the scale, so do the objectives in each scale.

This view of an energy system is possible due to concepts well-known in the field of distributed systems theory. Distributed systems operate from the bottom-up, which is in tune with our approach to an energy network. The idea of taking small units and making them cooperate according to established metrics allows us to define goals and objectives for a network at different scales. The resulting picture is of a network of networks, for example:

Single household ⟶ Local area ⟶ Wider area ⟶ National grid

Our vision of an SLES is that, through a distributed approach, the elements composing a single household behave in unison through cooperation. The consequence of this is that when analysing a network of households, each house can be seen as a single entity. This method can be taken forward until we reach the national grid level. An advantage of this approach is the clear determination of roles in a particular setting, i.e., the scope and objectives are clear for each layer.

# The layered view of SLES

As more elements are added to the network, describing a network in terms of scales and layers allows us to tackle the increasing complexity of an SLES.

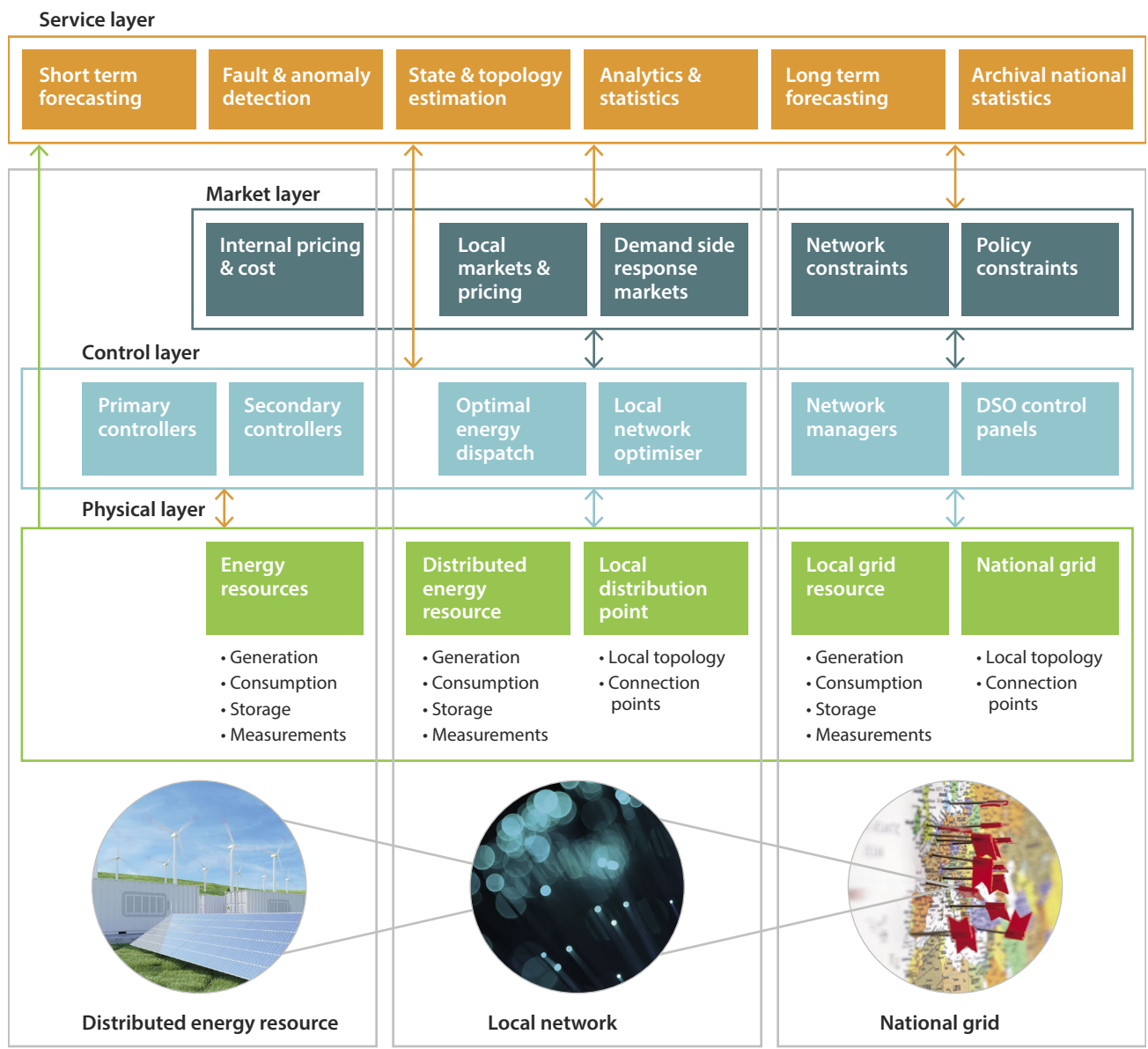

**Figure 1**: Abstraction layers and their components in digitised energy systems: DER, Local network and National Grid scales

www.energyrev.org.uk

Figure 1 shows the relationship between layers at different system scales and respective layer components in digitised energy systems. One of the most important features of this conceptualisation is the access to service layers for each of the featured scales, which contain crucial operations that facilitate the operation of both control and market features. The abstracted view of the various components and their connections can be seen in Figure 2, where most specifics such as technologies and individual goals of market and control components are removed, while a non-exhaustive list of services is provided for clarity.

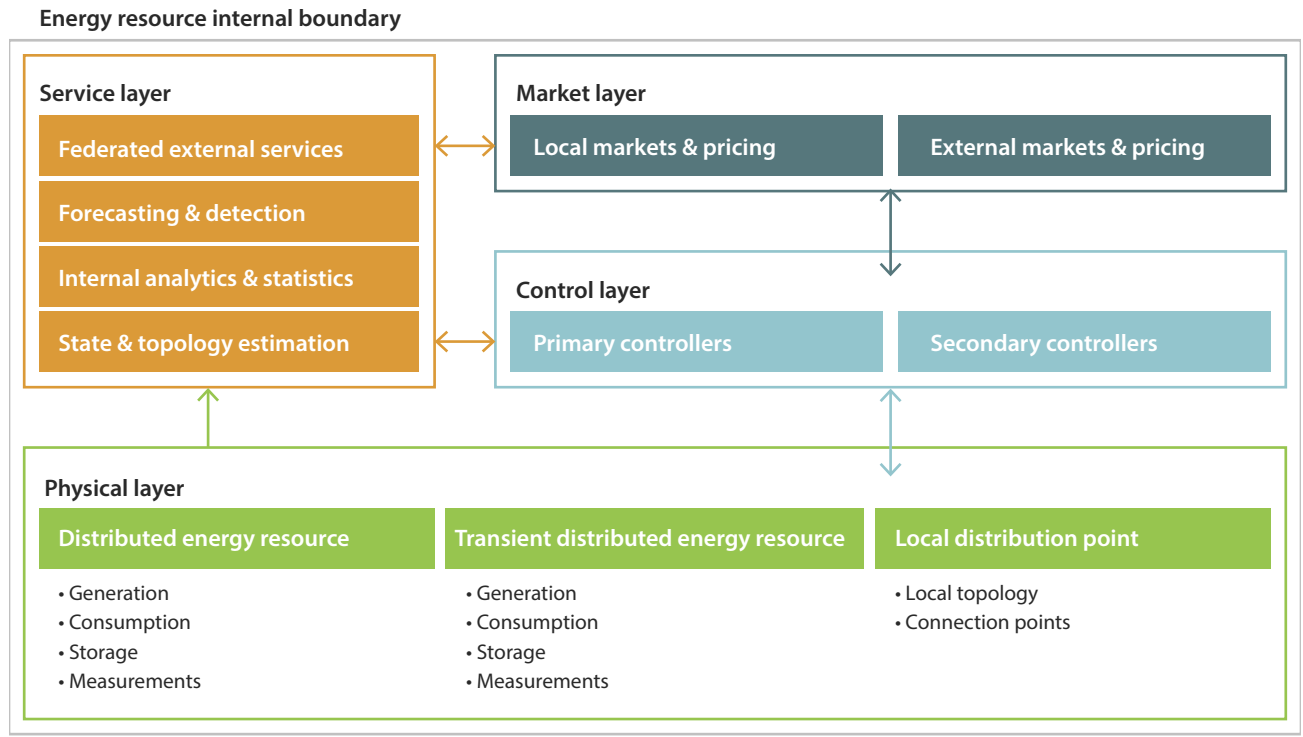

**Figure 2**: Energy Resource Internal Boundary: Abstraction layers and their components in digitised energy systems

**Physical Layers** are the physical and communication assets in the network. They are responsible for transmitting data to other components and enacting commands. The physical layer is broken down in a scalable way so that these components can be found on local, regional and national level. On a local level a DER resource may be a solar panel, battery or consumer, while a transient DER can be an electric vehicle and the distribution points its connection to the grid. On a regional level the DER resource is replaced by the connection point of the contained smaller system and distribution points are connections to the medium or high capacity lines.

**Control Layers** are the decision-making part of the systems. Layered controllers make decisions and optimise set points based on market and service information. These controllers are envisioned as an extension of the tertiary control where optimisers and higher level controllers calculate and provide set-points for lower level controllers and other set-points components in the architecture.

## Layered control and optimisation

The need for optimising and controlling the energy system in line with its objectives demands a layered control structure. This architecture can be separated into three layers: primary, secondary and tertiary control layer (see Fig. 3).

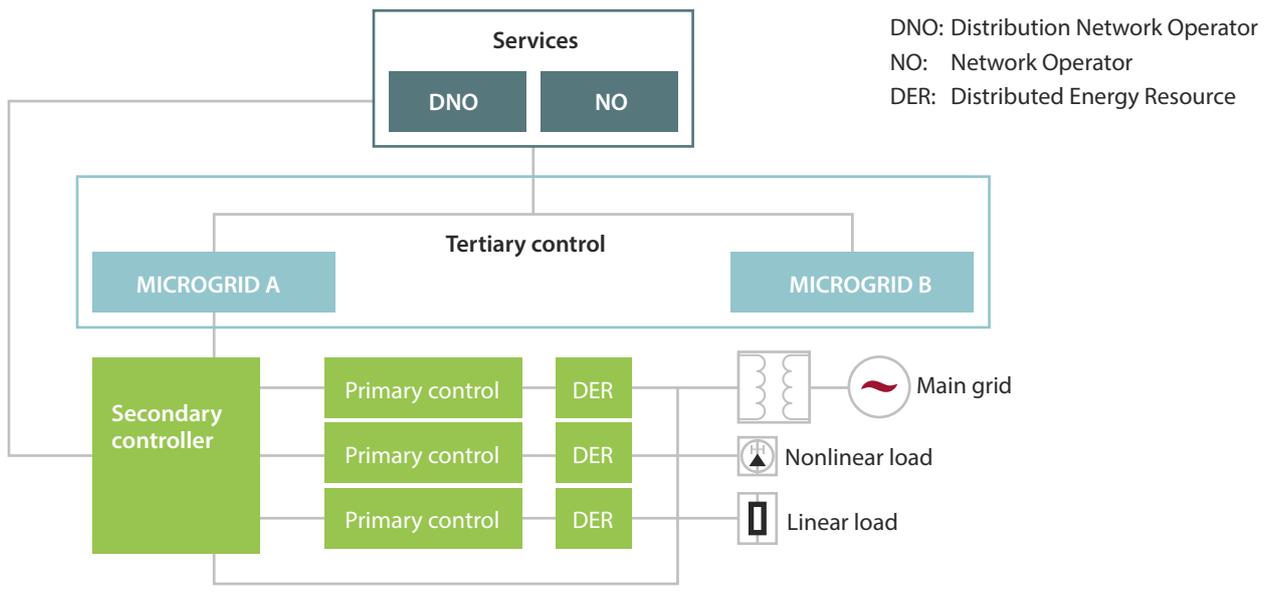

**Figure 3**: Controller Layers and interconnection in a DER abstraction

The first layer incorporates local optimisers with given costs and local constraints. It stabilises the voltage and frequency in the network and provides plug-and-play capability for DERs. The secondary control layer is generally designed to produce a slower dynamic response over the primary, optimising local constraints and references, such as restoring the voltage and frequency, and correcting the deviations caused by the first layer. Tertiary control is responsible for economic dispatch or tariff strategies, active network power management and macro-optimisation of the energy system operation.

The layered control approach could reduce the operational costs of the energy system while maintaining constraints and improving robustness. It could also be re-engineered to improve other objectives such as environmental and social factors.

**Market Layers** provide the controllers with costs and other requirements based on data from the service layer. The market component enables the trading of resources, the rewarding of certain behaviours and the satisfaction of grid constraints. At a small scale the market layer only contains the negotiated costs for energy, while at regional and national level the market layer is a selection of contract, P2P negotiation and various policy constraint components that all interact to allow local and regional components to set the price of energy and the constraints of its use.

**Service Layers** receive data from the physical, control and market layers of the energy system and store, process and archive that data. This layer takes the received data and produces required information about past, current and future performance, health and behaviour. The scale of the services can grow from single system forecasts and metrics to regional forecasting, topology analysis and system identification.

An example deployment would see services offered to small scale local energy systems on measurements and system identification, together with short-term forecasting. These services are crucial for the operation of the control and physical layers because they depend on these measurements and estimation to compute suitable control actions. Forecasting is also useful to a market layer to decide whether to store or transact available energy. At the next abstraction level, and owing to the distributed nature of the control and market layers, the service layers offer topology estimation services, more advanced forecasting, and a compendium of analytics. The implementation of such features involves wrapping new functionality around existing layers.

For example an optimisation-based energy dispatch controller computes suitable generation targets based on an optimisation cost; the market layer can modify this optimisation cost which would, in turn, affect how energy is dispatched. These modifications of the controller optimisation costs allow the market layer to define types of local pricing and thereby implement various demand-side management techniques based on need.

## Security layer on top of smart entities

The efficient and secure operation of SLES will be heavily reliant upon the secured communication system. Communication systems must transfer accurate and authentic data and information between different layers, enabling the SLES to maximise the utilisation of resources and benefits in a timely manner. However, the increasing presence of IoT technologies in the ICT infrastructure will also create new privacy or security challenges. These might include impersonation, data manipulation, eavesdropping, privacy breaches, disputes, and denial-of-service. For this reason, a comprehensive cybersecurity system should be designed to cover the following:

- **Entity Authentication** to ensure that entities can be assured of the identity of their communication partners in order to counter impersonation attacks. A challenge-response protocol can be adopted so that eavesdropped responses can be excluded.

- **Message Authenticity** to ensure the delivery of authentic messages to message receivers without any tampering or modification during transit. Digital signatures or message authentication code can be used to protect message authenticity.

- **Authorisation** to control access so that only authorised parties can get access and make relevant modifications within the system. It is usually coupled with Entity Authentication to protect systems from elevation of privilege attacks.

- **Confidentiality** to ensure that only the intended message receivers can read the message. Either symmetric or asymmetric encryption is widely used. It is usually combined with message authentication.

- **Users' Privacy Protection** to protect user privacy as much as possible based on the rule of "Principle of Least Privilege".

- **Availability** to ensure that a system or a system component is accessible when authorised entities require access. A combination of techniques, including attack detection, message categorisation and filtering techniques, as well as balancing load and resources, is essential to tackle denial of service attacks.

# Utility and future use

### Speeding up AI and Multi-Agent Systems adoption

A key step in enabling the wider adoption of AI and Multi-Agent Systems (MAS) is to characterise existing use-cases and define the boundaries of their various components. By doing so we enable the use of existing cyber-physical infrastructure to gather data and analytics on component behaviour and identify, quantify and validate the benefits brought by AI and MAS. The availability of quantified value and use-cases removes barriers around these new technologies and is key in gaining trust and developing explainability metrics for AI adoption.

The flexibilities built in by the communication network that links the independent components together also enables these components to be partially or fully replaced by more advanced AI and MAS counterparts, as technology evolves, while enabling the development of virtualised environments for robustness and integration testing.

At a local level agent-based abstractions can replace or upgrade existing controls and market components while also managing interactions with human and external agents. These local agents can be programmed to trade freely in P2P trading environments with local counterparts and national entities. This allows agents to be fast-tracked by following a well-defined set of rules, abstraction and permissions.

At the local authority level of a national grid, agent-based markets and negotiation clusters can replace traditional load balancing, network operations and contract-based systems by taking over the defined roles and endpoints of one or more of these services. More importantly, a distributed agent-based approach to SLES management provides an opportunity to optimise network control for social factors. These include fairer distribution of energy and better use of limited resources in the face of climate challenges.

# Conclusions

Energy systems of the future need to be broken down into independent and individual services, not just to democratise energy in a P2P environment but also to ease the development of competing products and services. The definition of these independent services, their roles and the value they offer can aid in the development and integration of AI approaches and the replacement of the existing communication architecture with dynamic P2P and agent-based channels.

Markets tailored to local needs coupled with individual and local level optimisation can aid in meeting environmental goals, reduce energy costs and make energy equitable, but only if we enable the digital components of energy systems to be updated, replaced and interchanged as technologies, needs and services change. These markets need to adapt throughout their lifecycle to integrate national policy requirements such as $CO_2$ emissions and local needs for fairness.

As the standardisation of industrial components sped up the development during the industrial revolution, defining and standardising the energy system component groups by roles and responsibilities can be a first step into fast-tracking their development towards fully digitised, flexibly interconnected, multi-layer plug and play architectures for energy systems.

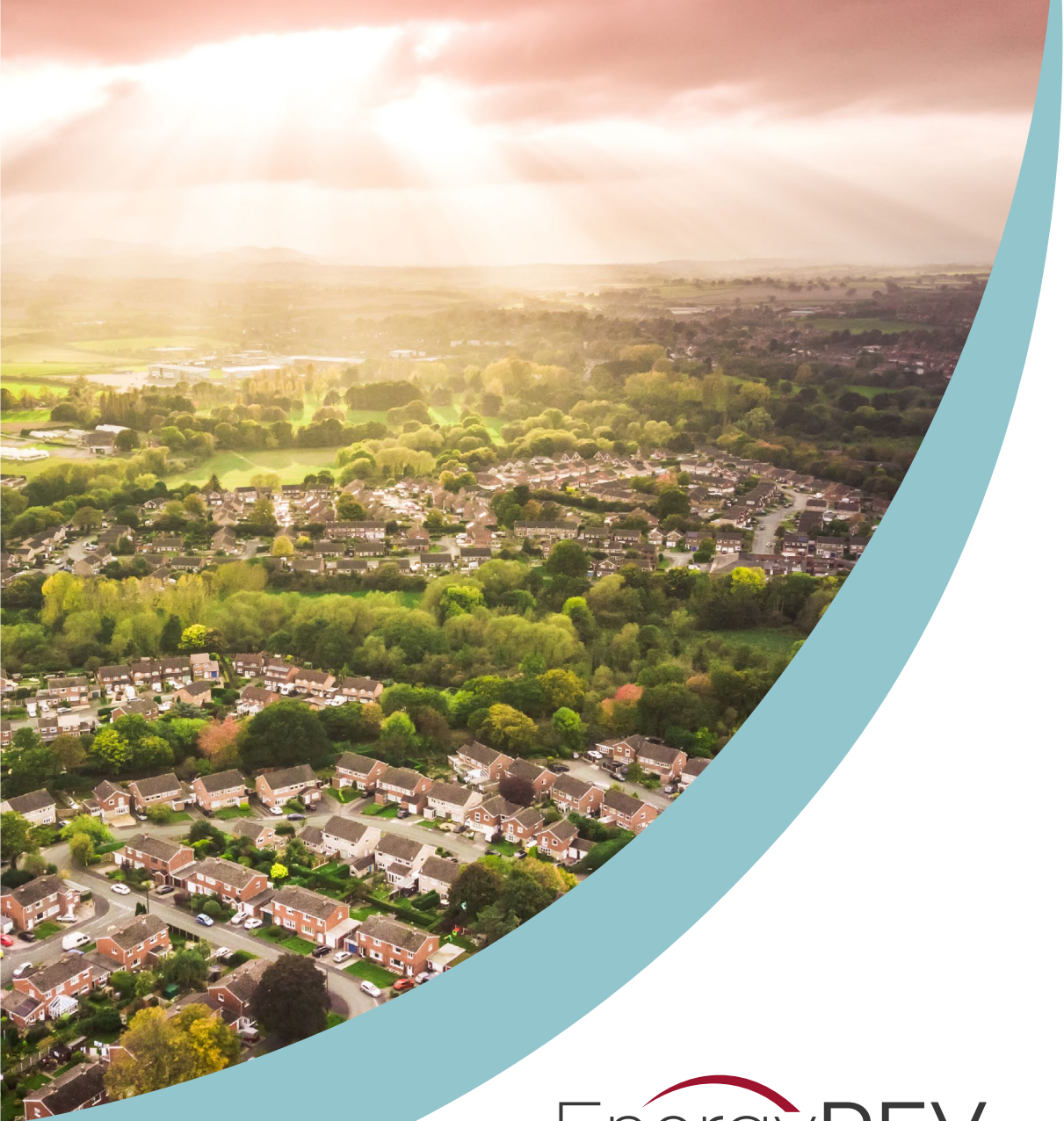

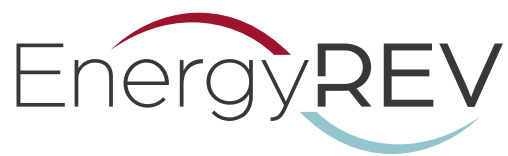

## Want to know more?

🌐 www.energyrev.org.uk

🐦 @EnergyREV_UK

in EnergyREV

✉ info@energyrev.org.uk

Sign up to receive our newsletter and keep up to date with our research, or get in touch directly by emailing info@energyrev.org.uk

### About EnergyREV

EnergyREV was established in 2018 (December) under the UK's Industrial Strategy Challenge Fund Prospering from the Energy Revolution programme. It brings together a team of over 50 people across 22 UK universities to help drive forward research and innovation in Smart Local Energy Systems.

EnergyREV is funded by UK Research and Innovation, grant number EP/S031898/1

ISBN 978-1-909522-94-7

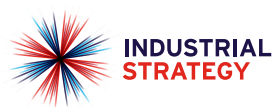 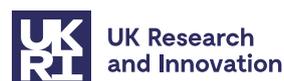